\documentclass[conference]{IEEEtran}
\IEEEoverridecommandlockouts
\usepackage{cite}
\usepackage{amsmath,amssymb,amsfonts}
\usepackage{amsthm}
\usepackage{algorithm}
\usepackage{algpseudocode}
\algrenewcommand\algorithmicrequire{\textbf{Input:}}
\algrenewcommand\algorithmicensure{\textbf{Output:}}
\usepackage{graphicx}
\usepackage{booktabs}
\usepackage{xcolor}
\usepackage{tikz}
\usetikzlibrary{arrows.meta,positioning,fit,backgrounds,calc}
\usepackage{microtype}
\usepackage[hidelinks]{hyperref}
\usepackage{url}

\newtheorem{theorem}{Theorem}

\newtheorem{definition}{Definition}

\newcommand{\obs}{\mathit{obs}}
\newcommand{\run}{\mathit{run}}
\newcommand{\Lvideo}{L_{\mathrm{video}}}
\newcommand{\eqL}{\approx_{L}}
\newcommand{\eqv}{\approx_{\Lvideo}}
\newcommand{\tool}{\textsc{Vid2Prog}}
\newcommand{\sbt}{\texttt{.sb3}}

\begin{document}

\title{Checked Program Recovery from Execution Video:\\ A Sound Oracle for Untrusted Generators}

\author{
\IEEEauthorblockN{Yuan Si and Jialu Zhang$^{*}$\thanks{Corresponding author: Jialu Zhang.}}\\
\IEEEauthorblockA{University of Waterloo, Waterloo, Canada\\
\texttt{yuan.si@uwaterloo.ca}, \texttt{jialu.zhang@uwaterloo.ca}}
}

\maketitle

\begin{abstract}
A growing class of tools recovers a program from observations of its behavior using an untrusted generator, a neural model or a search, that proposes candidates with no correctness guarantee. We study how to make such recovery trustworthy, in the concrete setting of recovering a runnable Scratch program from a recording of its execution. The recording shows what the program does but never its code; many programs produce the same video, so the source cannot be recovered, and the right target is a program that behaves the same as far as the camera can tell, made precise with a lens. The core is a two-tier validation oracle with a deliberate verdict asymmetry. A static checker proves lens-equivalence to a reference and issues a certificate that, granting the partial-order independence quotient adequate, never accepts a wrong program; a renderer can only refute or witness finite agreement, never certify. Around it, \tool{} reads each sprite's motion, visibility, and timing from the video and a known-asset manifest and synthesizes a candidate source-free; a closed loop renders and runs recovery again for ground truth. Under the exact lens the oracle makes no false accept on 246 labeled differing pairs, including an adversarial battery built to trap its concurrency quotient; on inputs outside the vocabulary and on real projects it abstains or refutes, accepting none we test. In-vocabulary recoveries reproduce their source frame for frame and 80\% earn a static certificate, while whole real projects, mostly outside the vocabulary, recover at 14\%, a vocabulary-bound rate the system never inflates with a wrong answer. A frontier vision-language model recovers none of the matched programs single-shot, which oracle-in-the-loop repair lifts only to a few while the structured pipeline recovers all, the gap a sound checker makes for an untrusted generator.
\end{abstract}

\section{Introduction}
\label{sec:intro}
A twenty-second screen recording of a Scratch game carries enough information for a person to grasp what the program does, and a neural model or a structured search can be asked to turn that recording back into a runnable \sbt{} file a learner could open and edit. The code never appears on screen; what the camera captures is the rendered output of an execution, sprites moving, costumes cycling, the stage redrawing thirty times a second. Reconstructing an executable program from this signal is the problem we study, and the difficulty that organizes the paper is not only producing a candidate but trusting it: a recovered program is the output of an untrusted generator that perceives and guesses, and matching a handful of frames is no proof it behaves like the original. Our answer is a sound checker that certifies a candidate or refuses it, and is never allowed to do the reverse.

Scratch is among the largest novice programming communities, with over a hundred million publicly shared projects, and learners produce far more recordings of running projects than they keep as editable source. A recording whose source is lost cannot be analyzed, tested, tutored, or remixed; recovering an editable program reopens it to every tool that reads source.

This problem sits apart from the inverse problems software engineering knows. Binary decompilation reads instructions that encode the computation directly; here the instructions are gone and only their visible consequences remain, so control flow, internal state, and object identity are inferred from how appearance changes over time.

The central difficulty is that many distinct programs produce the same video. A variable the program never renders, a branch the execution never takes, two independent scripts in either order: none of these leaves a trace a camera can see. Asking to recover the original source is therefore ill-posed. The question that is well-posed is whether a recovered program reproduces what was seen and behaves like the original on everything the camera could observe. We make this precise by treating a video as one realized point of a lossy observation map parameterized by a lens, a description of which execution facts are visible. The camera fixes the lens. The strongest claim any method can earn is observational equivalence under that lens, and a recovered candidate is validated under that lens by a two-tier oracle whose static tier alone mints an equivalence certificate.

Two commitments shape the system. Equivalence is certified, never assumed: a static checker re-derives a candidate's behavior from its \sbt{} bytes and mints a certificate only when it proves equivalence to a reference, available wherever a reference exists, while a video alone yields the weaker render witness; a renderer can expose a divergence but never mint a positive verdict, because matching finitely many frames cannot establish agreement on the infinitely many executions a camera never witnessed. And the part a camera cannot recover is stated as a theorem: we construct five families of programs pairwise indistinguishable under the camera lens, a constructive lower bound on the non-recoverable residual.

\tool{} is the untrusted generator these commitments guard: it perceives sprite tracks from pixels, fits structural motifs that explain the tracks, synthesizes a candidate program, and routes it through the validation oracle. The pipeline recovers the motion, visibility, and frame-counter control of a program; the sprite cast, costumes, and backdrop are supplied as a known-asset manifest, which is the lens under which recovery is posed. When passive observation is insufficient, because a behavior is gated on an input that never occurred, the system queries the original program as a black box, synthesizes the missing input, elicits the hidden behavior, and recovers the input condition. The forward renderer that produces videos also serves as a labeled-data generator and as the refutation engine, which yields a closed loop: render a program, recover the program from the video, check the result. The recovery rate is then the metric and the task definition at once, and the benchmark needs no human annotation.

The oracle never accepts a wrong program: no false accept on 246 labeled differing pairs or on an adversarial battery built to trap its concurrency quotient, and outside the vocabulary and on real projects it abstains or refutes. Within the vocabulary the generator is strong, every recovery reproducing its source frame for frame and 80\% earning a static certificate, both stable across ten seeds; whole real projects, mostly outside the vocabulary, recover at 14\%, the vocabulary-bound rate paid in full. Recovery holds from thirty frames per second down to five and through a second renderer's compressed video, an active loop recovers key-gated behavior, and a frontier vision-language model recovers none of the matched programs from frames, a perception gap the pipeline's least-squares fit closes, while the oracle catches its errors and the pipeline recovers all. The in-vocabulary corpus fixes the synthesis ceiling and makes perception the variable under test.

This paper makes four contributions.
\begin{itemize}
\item A two-tier validation oracle that makes untrusted generative program recovery trustworthy: by a verdict asymmetry only its static tier mints a lens-equivalence certificate, sound relative to the compiled IR semantics and the stated independence-quotient adequacy assumption, while a renderer only refutes or witnesses finite agreement, never certifies, yielding zero false accepts over 246 labeled different pairs (Section~\ref{sec:oracle}).
\item A formalization of video-based recovery as inversion of a lens-parameterized observation map, with an identifiability lower bound characterizing what a camera cannot recover (Sections~\ref{sec:formal} and~\ref{sec:theory}).
\item \tool, a source-free neuro-symbolic pipeline that recovers a program's motion, visibility, and frame-counter control from rendered pixels under a known-asset lens, through verification-based motif synthesis and a black-box active-inference loop for input-gated behavior (Section~\ref{sec:pipeline}).
\item A closed-loop, label-free benchmark and an evaluation of soundness, recovery, discrimination, robustness, and a vision-language baseline, reported with confidence intervals (Section~\ref{sec:eval}).
\end{itemize}
The static tier implements a published lens-equivalence algorithm~\cite{scratchlens}; the contribution is the soundness discipline around it, the two-tier oracle and its verdict asymmetry, the lens formalization and identifiability bound, the pipeline, the closed-loop benchmark, and FastRender, a deterministic renderer used for cross-renderer validation.

\section{Overview}
\label{sec:overview}
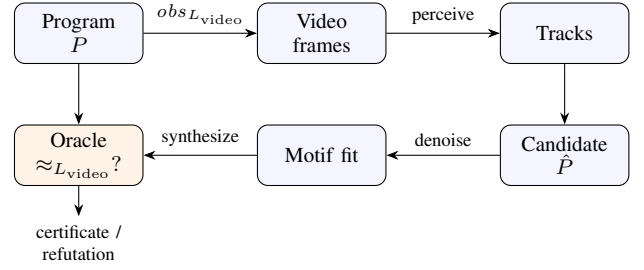
\begin{figure}[t]
\centering
\begin{tikzpicture}[
  font=\footnotesize,
  box/.style={draw, rounded corners, inner sep=2pt, minimum height=8mm, minimum width=17mm, align=center, fill=blue!4},
  obox/.style={draw, rounded corners, inner sep=2pt, minimum height=8mm, minimum width=17mm, align=center, fill=orange!10},
  >={Stealth[length=1.6mm]},
  node distance=8mm and 15mm]
\node[box] (prog) {Program\\ $P$};
\node[box, right=of prog] (vid) {Video\\ frames};
\node[box, right=of vid] (perc) {Tracks};
\node[box, below=of perc] (syn) {Candidate\\ $\hat P$};
\node[box, left=of syn] (den) {Motif fit};
\node[obox, left=of den] (orac) {Oracle\\ $\eqv?$};
\draw[->] (prog) -- node[above,font=\scriptsize]{$\obs_{\Lvideo}$} (vid);
\draw[->] (vid) -- node[above,font=\scriptsize]{perceive} (perc);
\draw[->] (perc) -- (syn);
\draw[->] (syn) -- node[above,font=\scriptsize]{denoise} (den);
\draw[->] (den) -- node[above,font=\scriptsize]{synthesize} (orac);
\draw[->] (prog.south) -- (orac.north);
\node[font=\scriptsize, align=center, below=4mm of orac] (verdict) {certificate /\\ refutation};
\draw[->] (orac.south) -- (verdict.north);
\end{tikzpicture}
\caption{From program to video and back. Perception and synthesis propose a candidate from the video and asset manifest alone. The validation oracle, which reads the source withheld from recovery, then certifies statically where a reference is available, witnesses by replay against the frames otherwise, or refutes.}
\label{fig:pipeline}
\end{figure}

Figure~\ref{fig:pipeline} traces one video through the system. A Scratch program drives a sprite rightward and hides it after a fixed number of frames. The renderer produces a sequence of frames, the only input the recovery sees.

Perception reads each frame, segments the sprite from the background, and reports its centroid and visibility. The result is a noisy track: the measured position drifts by a few pixels per frame around the true trajectory. A least-squares fit over the whole track absorbs this noise and returns an exact per-frame displacement, which is why recovery from imperfect pixels is feasible. Synthesis searches for a small program whose forward simulation reproduces the denoised track, and it accepts a motif only when the simulation matches the track exactly. The candidate is a runnable \sbt{} file.

In the benchmark the validation oracle decides acceptance: it compiles the candidate and the withheld source and compares them under the camera lens. Here a subtlety appears that the rest of the paper builds on. Suppose the original keeps a private counter it never displays. A recovered program without that counter renders an identical video. The two are equivalent under the camera lens, because no camera could separate them, and the oracle accepts the smaller program. Acceptance means equivalence on what the lens observes, and the facts the lens hides are exactly the residual our identifiability result quantifies.

This example is why the target is a class, not a program: a method insisting on the original source would have to recover the counter, impossible from the video, while one targeting the class succeeds whenever it returns any member the oracle accepts. The shift from a point to a class makes the problem well-posed, and the camera forces it.

\section{Problem Formalization}
\label{sec:formal}
We fix the space $\mathcal{P}$ of loadable Scratch programs. Execution is a relation rather than a function, because the outcome depends on exogenous choices: an input timeline $\iota$ that lists green-flag, key, mouse, and broadcast events, a random seed $r$ that pins the generator and the clock, and a legal schedule $s$ that orders concurrent scripts and clones at tick boundaries. Treating execution as a relation places the schedule on the same footing as the input and the seed, which is why script order is a non-identifiability family, and it forces the acceptance criterion to quantify over schedules. Once the three choices are fixed the trace is determined, written $\run(P,\iota,r,s)$, and all nondeterminism lives in the choice of the triple, not inside the step relation.

A lens $L$ names which facts of a trace are observable. Lenses carry an order: $L \sqsubseteq L'$ means $L'$ observes at least as much as $L$, and the two relate by a projection $\pi$ with $\obs_L = \pi \circ \obs_{L'}$. A video is one realized observation under the camera lens
\[
\Lvideo:\ \text{per sprite per frame, } (x,y,\text{dir},\text{costume},\text{visible},\text{size}),
\]
together with rendered monitor values when shown. Audio, pen-buffer layout, and internal state fall outside it. On this order the camera lens sits between a final-state lens that sees only the last configuration, which is coarser, and a complete lens that sees every internal effect, which is finer. Fixing the camera lens turns an open-ended request to recover the program into a precise request to recover a member of one equivalence class.

The order on lenses is realized by a projection we construct explicitly. The checker represents an observation as a multiset of feature tokens, each tagged by the lens axis that emits it. To coarsen from $L'$ to $L$, $\pi$ deletes every token whose axis is on in $L'$ and off in $L$. The factorization $\obs_L=\pi\circ\obs_{L'}$ then holds by construction, and the monotonicity it supports does not rest on any per-axis guard scattered through the checker. This is the one place we build the structure the theory needs, so the recoverability frontier of Section~\ref{sec:theory} is a property of the construction, not a conjecture about the implementation.

Two programs are equivalent under $L$, written $P \eqL Q$, when they produce identical $L$-observations for every input, seed, and legal schedule, the lens-parametric equivalence introduced by~\cite{scratchlens} and decided by our oracle's static tier. The relation $\eqL$ is the kernel of the observation map, and the quotient $\mathcal{P}/\!\eqL$ is the codomain of recovery: recovery targets an equivalence class, not a point.

A video witnesses one execution, so acceptance is weaker than full equivalence and separates what was seen from what was not.

\begin{definition}[Reference-validated acceptance]
\label{def:accept}
A candidate $\hat P$ is accepted for a video $v=\obs_L(\run(P,\iota,r,s))$ when (i) $\obs_L(\run(\hat P,\iota,r,s))=v$, equivalence on the witnessed execution, and (ii) on every unwitnessed input, seed, and schedule, $\hat P$ exhibits no $L$-observable that $P$ could not, a refinement off-trace.
\end{definition}

Clause (i) is the trace witness, checkable from the video; clause (ii) names the source and is established only by the static tier against a reference. A render witness checks clause (i) over sampled frames alone, a trace acceptance short of the off-trace refinement (ii) demands, so the two tiers earn different halves of the definition. By translation validation, narrowing open nondeterminism is sound while adding an unproduced observable is not, so a static certificate gives equivalence on what was seen and at least the refinement clause (ii) asks off-trace; being a symmetric multiset equality it in fact delivers full $\eqL$, more than (ii) demands, though never source identity.

\section{What a Camera Cannot Recover}
\label{sec:theory}
The observation map is not injective, and the ways it collapses distinct programs are the structural reason recovery targets a class. We make the collapse constructive.

\begin{theorem}[Non-identifiability, lower bound]
\label{thm:t1}
The kernel $\eqv$ contains at least five effectively constructible, individually non-empty families of pairs $(P,Q)$ with $P\neq Q$ and $P\eqv Q$: (i) never-rendered state, (ii) dead or effect-masked code, (iii) reordered independent statements, (iv) reordered independent scripts, and (v) externally observed latent state. The five families are pairwise distinct (Table~\ref{tab:axes}).
\end{theorem}

Each family carries a generator that turns any program into a distinct, camera-equivalent variant, with the equivalence witnessed by a static proof for three families and by a render for the two intrinsic ones. The first family inserts a variable declared but referenced by no block; it produces no transaction and no observable feature, so the feature multisets of the two programs are equal and the checker proves equivalence statically. The second wraps a real block stack under a constant-false guard, which the compiler folds away before comparison, or leaves a stack with no hat, which the compiler drops; the surviving features coincide. The third reorders two writes to distinct invisible variables, which the partial-order normal form maps to the same representative. The fourth swaps two independent scripts, which become unordered facts that project identically under the camera lens. The fifth writes a cloud variable the program never displays, which the lens projects to nothing. Table~\ref{tab:axes} lists the families with the certificate that discharges each and the finer lens, where one exists, at which the difference becomes observable.

The boundary of each generator is sharper than the family name suggests, and we state it precisely. A variable that is written and then read leaves a feature even when the camera never renders it, so it is a kernel member the static checker cannot certify, and its non-emptiness is witnessed by a render, not a static proof; code after an unconditional stop, dead yet still compiled, routes to the render witness for the same reason. The split is by certificate, not by family: some collapses carry a static proof, and some are real but witnessable only by replaying both programs, which grants no soundness credit and is reported as such.

The statement is a lower bound, and we mark its boundary. Two families, never-rendered state and external latent state, are invisible at every lens: the difference is the absence of a feature, which no checker can positively mint, so they admit no refuter. The other three become observable at a finer lens, which makes them constructive in both directions. We do not claim the five families exhaust the kernel; claiming exhaustion would require a normal-form theorem, that every camera-equivalent pair is connected by moves drawn from these five families, which is likely false. A lower bound is the right form of the claim, and it still delivers what the system needs, a constructive account of the residual that recovery targets a class to absorb.

\begin{table}[t]
\centering
\caption{Five non-identifiability families, the certificate that proves camera-equivalence, and what separates each, a finer lens or intervention where one exists}
\label{tab:axes}
\footnotesize
\setlength{\tabcolsep}{4pt}
\begin{tabular}{@{}lll@{}}
\toprule
Family & Why a camera cannot see it & What separates it \\
\midrule
never-rendered state & no block reads the variable & none \\
dead / masked code & folded by guard pruning & flip guard \\
statement order & shared normal form & step-trace lens \\
script order & disjoint, unordered facts & causal lens \\
external latent state & no external resource shown & none \\
\bottomrule
\end{tabular}
\end{table}

A recovery returns one program, and the class it represents is infinite, so the system has to choose a member. We fix the choice with a description-length prior. Among the loadable programs in a class, order by code size first and break ties by a canonical serialization, with a penalty on any block the lens cannot see. The penalty pushes the choice toward a program with no camera-invisible structure, a counter that is never displayed, for example, is dropped from the recovered program.

\begin{theorem}[Representative]
\label{thm:t2}
For a lens $L$ and this prior, the least-cost member of a class exists and is unique. Within the static tier's canonical fragment, a static certificate witnesses it equivalent to the source under $L$ when the source leaves no $L$-invisible feature the representative drops; for a source that drops one, a written-then-read but unrendered counter for instance, or for a pair outside that fragment, the recovery may remain render-witnessed only.
\end{theorem}

Existence holds because the cost is integer-valued and the integers are well-ordered, so a least cost is attained without any compactness of the class. Uniqueness needs the tie-break well-founded and injective on programs: a cost level can hold infinitely many programs, and a merely total order on an infinite set need not have a least element, while length-then-lexicographic order on finite serializations does, and provided the canonical serialization is injective on programs, a property we require of the \sbt{} canonical form rather than prove, the least string names one program. The representative is the smallest program under this prior; a different prior selects a different member, so we never call it canonical absolutely. The target is then well-defined and certifiable in the supported fragment whenever the source hides no such feature.

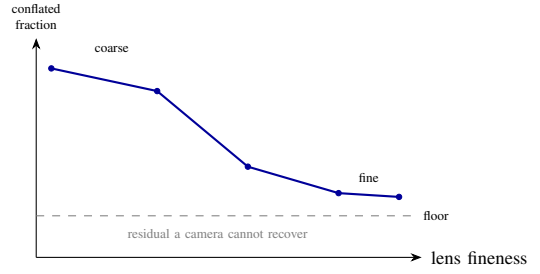
\begin{figure}[t]
\centering
\begin{tikzpicture}[font=\scriptsize,>={Stealth[length=1.5mm]}]
\draw[->] (0,0) -- (5.1,0) node[right]{lens fineness};
\draw[->] (0,0) -- (0,2.9) node[above,align=center,font=\tiny]{conflated\\ fraction};
\draw[thick,blue!60!black] (0.2,2.5) -- (1.6,2.2) -- (2.8,1.2) -- (4.0,0.85) -- (4.8,0.8);
\filldraw[blue!60!black] (0.2,2.5) circle (1pt) (1.6,2.2) circle (1pt) (2.8,1.2) circle (1pt) (4.0,0.85) circle (1pt) (4.8,0.8) circle (1pt);
\draw[dashed,gray] (0,0.55) -- (5,0.55) node[right,font=\tiny,black]{floor};
\node[font=\tiny,align=center] at (1.0,2.75) {coarse};
\node[font=\tiny,align=center] at (4.4,1.05) {fine};
\node[font=\tiny,align=center,gray] at (2.4,0.32) {residual a camera cannot recover};
\end{tikzpicture}
\caption{The recoverability frontier. The fraction of a corpus a lens conflates falls monotonically as the lens sharpens; the floor is the non-identifiable residual of Theorem~\ref{thm:t1}.}
\label{fig:frontier}
\end{figure}

The projection that relates lenses also gives a monotonicity we record as an internal check, not a result. If $L\sqsubseteq L'$ through a total projection $\pi$ with $\obs_L=\pi\circ\obs_{L'}$, then ${\approx_{L'}} \subseteq {\eqL}$: applying $\pi$ to both sides of an equality preserves it, needing neither injectivity nor lattice structure. So the fraction of a corpus that collapses to a common class is monotone non-increasing as the lens sharpens, and sweeping the lens from coarse to fine the curve cannot rise (Fig.~\ref{fig:frontier}). This holds by construction, so an inversion would signal a token-leak defect in the lens-coarsening code, and our corpus shows none. The curve traces the territory the camera gives up: the coarse end measures how much a camera conflates, the fine end how much a complete view separates.

\section{The Validation Oracle}
\label{sec:oracle}
Validation rests on a checker that tests equivalence under a lens and never errs toward acceptance. The oracle grades a candidate in the benchmark and in any paired or regression setting; it is not part of source-free deployment, where the render tier alone applies. The oracle has two tiers with deliberately different authority.

The first tier is static and needs two programs to compare. Following the algorithm of~\cite{scratchlens}, which we implement, it attempts to prove lens-equivalence and returns unknown when it cannot: it compiles both programs to a canonical intermediate form, builds an alpha-renamed feature multiset per trigger, and quotients same-trigger effects by a partial-order normal form~\cite{mazurkiewicz1987trace,flanagan2005dynamic} so that independent operations in either order compare equal. It mints a positive verdict only when the two multisets are equal, or when the abstract final-state signatures match under the final-state lens. Every field is re-derived from the compiled form of the candidate, and no value supplied by the producer enters the verdict. Any other outcome defaults to different or unknown, and unknown is never an acceptance.

\begin{theorem}[Static soundness, assumption-relative]
\label{thm:sound}
If the static tier reports equivalent for $P$ and $Q$ under lens $L$, and the partial-order independence quotient is adequate for the compiled semantics, then $P \eqL Q$.
\end{theorem}

Proof sketch: the checker emits lens-tagged transactions from the compiled IR, and equality of the per-trigger multisets after lens projection and alpha-renaming gives agreement on every $L$-observable effect. The one remaining step is the stated assumption, that the partial-order quotient preserves the execution-order independence of the compiled semantics (Section~\ref{sec:discussion}).

The second tier renders. It replays the candidate under the schedule that produced the video and compares its per-frame observations against the recorded frames, a check that needs no reference program and so applies even when the source is lost. This tier refutes by reporting the first frame and field at which the two diverge, and it reports agreement over the sampled frames, which we label probable equivalence. It does not mint a certificate. The reason is a quantifier gap.

\begin{theorem}[Verdict asymmetry]
\label{thm:t3}
A positive equivalence verdict is minted only by the static tier. A renderer in the loop refutes a candidate with a localized counterexample or reports agreement on finitely many frames; it never certifies equivalence.
\end{theorem}

Refutation is monotone: one divergent frame falsifies a universal claim. Confirmation is not: agreement on a finite sample says nothing about the unwitnessed executions, of which there are infinitely many. Granting the renderer certifying power would let a candidate that happens to match the sampled frames pass while differing everywhere else. The asymmetry keeps the renderer useful as a refuter and powerless as a certifier. Two conditions make the refuter sound. Replay is deterministic, achieved by a tick-derived clock and a pinned schedule, and the oracle abstains to unknown when determinism fails. Refutation replays under the schedule that produced the video, so a divergence caused by legal schedule reordering, family (iv) of Theorem~\ref{thm:t1}, is not mistaken for a real difference.

The split gives a quality ladder finer than a single accept bit. The bottom rung is over-fit replay: a program that hard-codes the witnessed poses, passes the renderer over the witnessed window, and diverges the moment it extends, so it carries no certificate. The middle rung fits the rule behind the motion; the synthesizer emits only such rule-form programs, never a pose table, so the doubled-window replay is a consistency check, not a memorization test. The top rung is a statically certified program, proved equivalent for every input and schedule. Reporting which rung a recovery reaches makes the soundness claim legible.

The static tier is sound but incomplete: on data-dependent control flow it returns unknown, not a proof, and a bounded search for a renaming that aligns two programs can leave a true equivalence unproven. Incompleteness is the right failure mode here, because unknown is never an acceptance and a missed proof costs coverage, not soundness. The error bar that matters is the unknown rate on programs that are equivalent yet syntactically distinct; it surfaces as the recoveries the static tier leaves render-witnessed rather than certified, the gap reported in Section~\ref{sec:eval}, not as a soundness failure. An identity round-trip, where a recovery equals its source, proves trivially, so it is not the measurement we report.

The soundness invariant ties the tiers together, relative to the compiled IR semantics and the partial-order independence relation the static tier quotients by, the one adequacy obligation we name in Section~\ref{sec:discussion}. A positive verdict is minted only by the static tier and only from the compiled bytes of the candidate; the renderer either refutes or abstains; any crash in the checker or a failed determinism guard fails closed to refutation. The recorded certificate is the lens, the two feature multisets, the alpha-renaming, and the equality verdict; a refuting render witness records the first divergent frame and field. Under this invariant a certified recovery is equivalent to its reference whenever the quotient of Theorem~\ref{thm:sound} is adequate, a static proof of equivalence a different program cannot produce, and Section~\ref{sec:eval} tests that adequacy directly with an adversarial battery that reorders dependent operations the quotient could wrongly commute. The render witness accepts a broader set, a recovery that reproduces the witnessed execution, and claims no more than that.

\section{The Recovery Pipeline}
\label{sec:pipeline}
\tool{} turns a video into a checked candidate in three stages: perception reads tracks from pixels, synthesis fits a generalizing program to the tracks, and the oracle of Section~\ref{sec:oracle} decides acceptance. A fourth stage, active inference, handles behavior that no passive video reveals.

\subsection{Perception}
Each frame is a $480\times360$ image. A foreground mask separates sprites from the stage, connected components give candidate blobs, and each blob yields a centroid and a bounding box. Identity across frames follows from a one-to-one assignment between detected blobs and the known cast~\cite{kuhn1955hungarian,bewley2016simple}, which keeps multiple sprites separated as long as they do not occlude. The position of a sprite is the centroid mapped to stage coordinates through a fixed offset between the centroid and the rotation center of the costume. We calibrate this offset once with a static grid sweep, which removes a constant bias and leaves the measured position exact to within a pixel across the central region of the stage.

The offset between the centroid and the rotation center is a property of the costume shape, and calibrating it matters more than its size suggests. A grid sweep that renders a sprite at known positions exposed a constant bias of more than ten pixels that a hand-set offset had left in place, hidden under the oracle tolerance yet far from exact; correcting it from the sweep left the perceived position exact to a pixel across the central stage, a defect the benchmark surfaced only because it checks every recovery against ground truth.

Two design choices bound what perception infers. The size, direction, and costume identity of a sprite are granted from the program under test, not inferred from a single blob, because a lone silhouette carries no reliable scale or orientation cue; a colorful or photographic backdrop is granted on the same footing and subtracted as a sprite-free render, which turns a non-white scene back into the white-stage case the mask expects. The boundary this draws is clean: the cast, costumes, sizes, directions, and backdrop form the asset manifest that fixes the lens, while position, visibility, timing, and the motion and frame-counter control behind them are what perception and synthesis recover from pixels, with event-gated behavior added by the active loop. The original source is neither supplied nor returned; the recovered program is a representative of its class.

\subsection{Verification-Based Synthesis}
Perception produces a noisy track, and synthesis must return a program, not a pose table. The vocabulary is a set of structural motifs: constant per-frame displacement, a single-segment glide that moves for a fixed number of frames and then stops, a single visibility transition that hides or shows the sprite at a frame, periodic costume cycling, and a clone family that spawns a fixed number of copies each running a shared behavior. A program is a loop over a frame counter whose body applies the fitted motifs.

Algorithm~\ref{alg:synth} fits the motifs and emits the program. Two ideas make it sound and precise. The first is a verification gate: the program built for a sprite is kept only when its forward simulation reproduces the track frame by frame (line~\ref{alg:gate}), which discards a motif that matches a prefix and then diverges and guarantees the emitted program reproduces what was seen. The second is the line fit \textsc{FitLine}, which turns noisy pixels into an exact integer rule. It takes the least-squares slope over the whole track, rejects the axis when the largest residual exceeds the perception tolerance, snaps a sub-perceptible slope to zero, and snaps a near-integer slope to that integer. A single frame carries several pixels of error, yet the slope of a line through sixty frames averages the error to a fraction of a pixel, so an integer-displacement program is recovered exactly and not merely to within tolerance. A glide the line fit rejects is recovered by splitting the steps into one contiguous moving run and a static remainder, and visibility and costume recover a single transition and a verified period in the same style.

\begin{algorithm}[t]
\caption{Verification-based motif synthesis}
\label{alg:synth}
\small
\begin{algorithmic}[1]
\Require track $T$ over frames $0..N$; tolerances $\tau,\zeta,\eta$
\Ensure program $\hat P$ with $\textsc{Sim}(\hat P,N){=}T$, else \textsc{Abstain}
\Function{Synthesize}{$T$}
  \For{sprite $s \in T$}
    \State $M \gets \emptyset$
    \For{axis $a \in \{x,y\}$}
      \State $d \gets \Call{FitLine}{s.a}$
      \If{$d \neq \bot$}
        \State $M \gets M \cup \{\,\textsc{changeBy}(a,d)\,\}$
      \Else
        \State $g \gets \Call{FitGlide}{s.a}$
        \If{$g = \bot$}\ \Return \textsc{Abstain} \EndIf
        \State $M \gets M \cup \{\,g\,\}$
      \EndIf
    \EndFor
    \State $M \gets M \cup \{\,\Call{FitVisibility}{s},\ \Call{FitCostume}{s}\,\}$
    \State $\hat P_s \gets \textbf{forever}\,[\,\mathit{tick}{+}{+};\ M\,]$
    \If{$\textsc{Sim}(\hat P_s,N) \neq T_s$} \label{alg:gate}
      \State \Return \textsc{Abstain} \Comment{verification gate}
    \EndIf
  \EndFor
  \State \Return $\Call{Assemble}{\{\hat P_s\}}$
\EndFunction
\Statex
\Function{FitLine}{$\mathbf{v}$}
  \State $(a,b) \gets \Call{LeastSquares}{\mathbf{v}}$
  \If{$\max_i |\mathbf{v}[i]-(b+a\,i)| > \tau$}
    \State \Return $\bot$ \Comment{not linear}
  \EndIf
  \If{$|a| < \zeta$}
    \State \Return $0$ \Comment{sub-perceptible}
  \ElsIf{$|a-\mathrm{round}(a)| < \eta$}
    \State \Return $\mathrm{round}(a)$ \Comment{integer}
  \EndIf
  \State \Return $a$
\EndFunction
\end{algorithmic}
\end{algorithm}

When no motif explains a track, synthesis abstains instead of emitting a program it cannot justify, which keeps a wrong answer off the table at the cost of coverage. The choice to abstain aligns the synthesizer with the oracle. A best partial guess would only hand the oracle a program to refute, and a confident point estimate would risk matching a misperception, which the refinement clause forbids; abstention turns both failure modes into a sound non-answer the benchmark reports in full. The cost is real: an abstention is a program the system could not recover, and we report it in the denominator, so the recovery rate is a rate over all attempts and not over a filtered subset.

Synthesis extends to clone families. A program that spawns a fixed number of clones, each running a shared behavior, appears in the trace as a growing set of identical sprites that share a name. Recovery reads the maximum clone count, the spawn interval, and the per-clone motion off the trace, and rebuilds the spawning loop and the clone body. The oracle compares clones by birth order, so a recovered family with the right count and the right per-clone behavior matches the source frame by frame. This closes the largest remaining gap in the vocabulary at the trace level, and we mark its boundary in Section~\ref{sec:discussion}.

\subsection{Active Inference for Input-Gated Behavior}
Active inference is an extension under a different access model, kept separate from passive recovery. A behavior gated on an input never appears in a passive recording. A sprite that moves only while a key is held stays still under a plain green-flag run, and no perception, however accurate, can recover a behavior that never occurs. Recovery here leaves the passive-video setting and takes black-box query access to the original program: the system supplies inputs and observes the rendered response, and never reads the source. This is the interactive setting of active learning, and its input differs: the original is a runnable build whose source is not in hand, a packaged or embedded build that runs but exposes no blocks.

Algorithm~\ref{alg:active} recovers the gate. It renders the program passively and, finding it static, holds each candidate key in turn and renders again, which is the synthesized experiment. A motion present only under a key reveals the gate, and the system emits ``forever, if the key is pressed, apply the motion.'' Confirmation is the careful step: the recovered program must agree with the original under both the passive timeline and the active one. Passive agreement alone is met by a do-nothing program, so the active timeline is the render witness that confirms the condition and rules out the vacuous answer.

\begin{algorithm}[t]
\caption{Active-inference gate recovery}
\label{alg:active}
\small
\begin{algorithmic}[1]
\Require renderable program $P$; candidate keys $K$; oracle \textsc{Eq}
\Ensure render-witnessed gated program $\hat P$, else \textsc{Abstain}
\Function{RecoverGate}{$P,K$}
  \State $T_0 \gets \Call{Render}{P, \langle\,\rangle}$ \Comment{no input}
  \If{$\Call{Moves}{T_0}$}
    \State \Return $\Call{Synthesize}{T_0}$ \Comment{not gated}
  \EndIf
  \For{key $k \in K$}
    \State $T_k \gets \Call{Render}{P, \textsc{hold}(k)}$ \Comment{synthesize input}
    \State $\delta \gets \Call{GatedDelta}{T_0, T_k}$ \Comment{motion only under $k$}
    \If{$\delta \neq 0$}
      \State $\hat P \gets \textbf{forever}\,[\,\textbf{if}\ \textsc{pressed}(k)\ \textbf{then}\ \textsc{apply}(\delta)\,]$
      \If{$\Call{Eq}{P,\hat P,\langle\,\rangle} \wedge \Call{Eq}{P,\hat P,\textsc{hold}(k)}$}
        \State \Return $\hat P$ \Comment{agrees on both timelines}
      \EndIf
    \EndIf
  \EndFor
  \State \Return \textsc{Abstain}
\EndFunction
\end{algorithmic}
\end{algorithm}

This loop is the one place where the contribution is irreducibly about acting on the program: the information about a gated behavior exists only in the program's response to an input it did not receive, and producing that response requires synthesizing the input. The choice of which input to try is an experiment-design question~\cite{settles2009active,angluin1987learning}, and a held key is the experiment that most separates a gated program from a do-nothing one. The territory is not marginal: a static census over real Scratch programs places roughly a third with input-gated visible behavior, structurally blind to passive observation, with an empty-program floor below ten percent, so the active loop's double-check against the vacuous answer matters.

Both algorithms share one soundness discipline. Synthesis emits a program only after its forward simulation reproduces the track, and active inference emits a gate only after the render witness confirms it under both timelines. Neither mints a certificate: the static lens-equivalence certificate is the only positive verdict the system mints, reserved for the static tier, which by the verdict asymmetry of Theorem~\ref{thm:t3} a renderer can never supply. What the algorithms produce is a candidate that reproduces the witnessed execution under the refinement clause of Definition~\ref{def:accept}; the static tier alone turns that into a certificate.

\section{The Closed-Loop Benchmark}
\label{sec:benchmark}
Evaluating recovery usually needs a labeled corpus of videos paired with their source programs, which is expensive to build and easy to bias. The forward renderer removes the need. A generator emits a program, the renderer turns it into frames, \tool{} recovers a candidate from the frames with no access to the source, and the validation oracle then compares the recovery to that withheld source under $\Lvideo$. The source is known because the loop produced it, so the ground truth is exact and free, and the recovery rate becomes both the metric and the definition of success. The source is an evaluation input only; recovery sees the frames and the asset manifest. An audit of the pipeline confirms the boundary: perception, denoising, and synthesis receive only the frames and the manifest, and the source reaches only the renderer that makes the video and the validation oracle.

The loop also defines a difficulty knob. Holding a program fixed and sharpening the lens raises the bar for a recovery, by the lens-frontier monotonicity, and perturbing a recovered program tests the other direction, whether the checker rejects a program that is close but wrong. We build near-misses by changing a single displacement by one or shifting a transition by a few frames, which probes the checker against the most adversarial inputs a recovery faces, programs that differ from the source by the smallest step the vocabulary allows.

Three guards keep the metric from inflating. A recovery has three outcomes, recovered, refuted, and abstained, and we report all three over the full denominator, so an abstention is never silently dropped or counted as a success. The metric is a sound checker, not a proxy such as pixel similarity, so a high rate cannot be bought by a lenient comparison. A do-nothing program, a tempting vacuous answer for a video with little motion, is ruled out by the near-miss test and the requirement that a recovery reproduce the witnessed observables: an empty program reproduces nothing and is refuted.

\section{Evaluation}
\label{sec:eval}
The corpus is procedurally generated inside the fragment the vocabulary covers, which fixes the synthesis ceiling and makes perception the variable under test; this isolates recovery-from-pixels from synthesis coverage, reported separately in RQ6. Each program is deterministic under a pinned seed and clock. The generator draws each program from constant, bounded-glide, and static motion along one axis, with small integer per-frame displacements, visibility and costume transitions at frames sampled over the clip, and one or two separable sprites; this space is broader than the canonical vocabulary of the synthesizer, so a non-canonical source must be recovered as a distinct lens-equivalent program rather than echoed, while a source already in canonical form is recovered exactly. The oracle runs at a position tolerance of 18 pixels for recovery from real pixels, which matches the perception noise floor, and at gold resolution for tests that probe the checker itself. Every rate carries a Wilson 95\% confidence interval~\cite{wilson1927probable}, the appropriate interval for proportions near one where a normal approximation understates uncertainty.

We ask six questions. RQ1: does the pipeline recover programs from rendered video? RQ2: does the oracle reject programs that are close but wrong? RQ3: does recovery survive degraded video? RQ4: can the active loop recover input-gated behavior? RQ5: how does a commercial vision-language model compare? RQ6: is the certifier sound, and how far does synthesis reach?

\paragraph{Perception Accuracy}
After offset calibration, the perceived position of a sprite matches the truth to within a pixel across the central stage, and the assignment step keeps two sprites' identities correct on every frame at the same accuracy. The remaining error is small, unbiased, and centered on the truth, the precondition the least-squares fit needs to average it away, so perception is not the bottleneck for in-vocabulary motion and the end-to-end rate reflects the vocabulary and the lens, not a perception ceiling.

\paragraph{Recovery From Video (RQ1)}
The first block of Table~\ref{tab:results} reports recovery under the render witness. Over 798 procedurally generated programs across ten seeds, every recovery reproduces its source frame for frame under the render witness: the recovered program replays in exact agreement with the original over the witnessed window and a window twice as long, with four refutations, no abstention, and no wrong answer, the three benchmark outcomes reported over the full denominator. Single-sprite and multi-sprite recovery each reach 397 of 399 ($99.5\%$, $[98.2,99.9]$), the four misses refuted and never accepted. The multi-sprite programs place two sprites in separate horizontal bands with independent motion and visibility, which the assignment step keeps apart for the full run. The recovery reproduces the observed trajectory exactly: the least-squares fit drives the per-frame displacement to integer precision, so the witnessed motion is matched frame by frame and not merely within tolerance, which establishes that the perception-to-synthesis path carries enough information to reconstruct these programs from pixels under the known-asset lens.

How strongly each recovery is certified is the finer question, and it separates the render witness from a static proof. By the verdict asymmetry of Theorem~\ref{thm:t3} the render witness refutes but never certifies, so a recovery it accepts sits on the middle rung of the quality ladder, a generalizing rule-form program; the over-fit bottom rung that freezes once the window extends is a baseline the synthesizer never emits, so the doubled-window agreement checks consistency, not memorization. The top rung is a static lens-equivalence certificate, and running the static tier on the recovered programs proves $80\%$ of them equivalent to their source at every lens (635 of 794 across ten seeds, single-sprite $87.4\%$ $[83.8,90.3]$, multi-sprite $72.5\%$ $[68.0,76.7]$). Treating each seed as the unit of analysis, the rate is $80.0\%$ with a seed-level $95\%$ interval of $[76.9,83.1]$ that brackets the pooled figure, so the headline tracks variance across corpora, not two lucky draws. The synthesizer earns the certificate by emitting canonical integer literals and initial poses, which makes an identifiable recovery byte-equal to its source. Without this canonicalization the same recoveries stay render-witnessed but only $15\%$ carry a static proof. This rate measures how exactly the recovery matches the source, a structural comparison placing every certified recovery within a pixel of it, not the checker discriminating syntactically distinct programs, the content of RQ2 and RQ6. The residual that no canonicalization reaches is the recoverability frontier in miniature: a recovery that drops camera-invisible structure, the minimal representative of Theorem~\ref{thm:t2}, or a bounded glide whose gating the witnessed trace underdetermines, matches its source frame for frame yet is statically distinct from a bulkier source. The system recovers the rule, not the trajectory. A full-vocabulary corpus with motion under occlusion makes this quantitative: 79 of 80 recover frame for frame but only 58 certify, the residual motion the occlusion makes non-identifiable.

\paragraph{The Oracle Discriminates (RQ2)}
A recovery rate is meaningful only if the checker rejects programs that are close but wrong. We build near-misses by perturbing each program minimally, a single displacement off by one or a transition shifted by a few frames, and ask the oracle to reject them. At gold resolution the oracle rejects all 33 genuine near-misses with zero false accepts, which shows the checker discriminates structurally. At the perception tolerance the rejection rate is 26 of 33. The seven near-misses not refuted at the 18-pixel tolerance deviate, over their whole trajectory, by less than that perception noise floor; under the exact camera lens, which sees position to the pixel, they are distinct, but the tolerant render witness the denoiser needs against real-pixel noise does not separate them. These seven are a perception-noise shortfall at the operating tolerance, not a soundness failure: the static certifier's soundness is defined under the exact lens, where the gold-resolution result rejects all 33, and the tolerance floor shrinks as perception sharpens. The synthesis vocabulary is a knob under an invariant oracle. Extending it to two-segment motion and arbitrary show/hide sequences recovers 40 programs of two-speed motion, bounce, and multi-transition visibility that the single-segment vocabulary left uncertified, now 40 recovered and 39 certified with zero false accepts, the base corpus unchanged; a harder suite of 30 three- and four-segment motions stays beyond reach, abstained on in full, none wrong.

\paragraph{Robustness (RQ3)}
Real recordings are compressed and slow. The second block of Table~\ref{tab:results} reports two stresses. Frame-rate decimation keeps every $k$-th frame, simulating capture from 30 down to 5 frames per second; the denoiser fits displacement against the true tick of each surviving frame, so a sparse sample recovers the same per-tick motion as a dense one. Motion recovery stays at 20 of 20 across every rate down to 5 frames per second, where a two-second clip is eleven frames. A straight line is fixed by any two points, so continuous motion costs nothing under decimation, while a discrete event such as a hide localizes only to within one sampling interval. A colorful, per-frame-noisy backdrop, handled by background subtraction, also recovers 20 of 20, which shows perception is not tied to a white stage. Separately, encoding the frames as H.264 across the full quality range leaves the measured position within five to eight pixels, so compression preserves the coarse spatial structure recovery depends on. The strongest test is a genuine recording: each program is rendered through FastRender, a separate deterministic renderer we contribute with its own rasterizer, and its H.264 video is decoded back to frames; perception, now reading pixels a different engine produced, recovers all 40 losslessly and 39 of 40 under visible compression (CRF~28), one abstaining and none wrong. These stresses bound the gap to genuine recordings from several sides, and continuous motion survives all of them because its information is spread across frames; the one stated limit is that at five frames per second a discrete event is pinned only to within a sixth of a second, the resolution the observation rate allows.

\paragraph{Active Inference (RQ4)}
The third block of Table~\ref{tab:results} reports the input-gated case. Each of 16 programs is static under a passive run and moves only while a key is held. \tool{} queries the program as a black box, synthesizes the held-key input, recovers the gate, and the oracle confirms that the recovered program reproduces both the passive and the active trace. All 16 are recovered, with passive and active agreement on every one. The double check matters: passive agreement alone is met by a do-nothing program, so requiring active agreement is what certifies the recovered condition. The static census we ran over 262 real Scratch programs places input-gated visible behavior at 31\%, the prevalence the active loop addresses when such a program reaches a viewer as a runnable build without source.

\paragraph{Vision-Language Baseline (RQ5)}
A natural question is whether a commercial model recovers programs directly from frames. We feed frames sampled at known ticks and the sprite list to a vision-language model and ask, at temperature zero, for each sprite's per-frame pixel displacement and any visibility transition as a JSON specification, build a runnable program from its answer through our reliable builder, and route the result through the same oracle. Table~\ref{tab:vlm} reports two models, the open-weight MiniMax-M3 and Claude Sonnet 4.6, with exact versions, run dates, and frame sampling specified in the evaluation protocol; we read them as evidence that an untrusted generator errs confidently, not as a tuned competitor. On a matched set of the 40 single-sprite programs the pipeline recovers in full, Claude Sonnet 4.6 recovers 0 and MiniMax-M3 recovers 1, the oracle refuting every wrong answer with no false accept. The model reads qualitative behavior such as a costume cycle but misreads the quantitative parameters, eyeballing a displacement where the pipeline fits the rule over every frame. The failure is perception, not synthesis: handed the perceived tracks as text, with no displacement left to eyeball, the same model recovers 29 of 40, and the oracle refutes the 10 it still gets wrong while accepting none. The pipeline's edge is the least-squares fit on noisy pixels; the oracle catches a wrong specification whoever produced it.

This is where the sound checker earns its place. Any generative recovery, a model or a search, produces candidates without a guarantee; the oracle converts them into trustworthy output by refuting the wrong ones and feeding the first divergence back for repair. We close that loop: the oracle returns to the model the first frame and field where the candidate's replay diverges from the recording, with the value the recording shows there, all observational, and the model revises. Over three rounds this lifts Claude from 0 to 3 of 40, concrete evidence that the refutations are actionable, while the structured pipeline with the same checker reaches all 40. The model proposes, the oracle disposes, and a search, neural or symbolic, repairs; the checker is what makes any of them trustworthy.

\paragraph{Soundness and Coverage (RQ6)}
Two measurements bound the system from the sound side. The certifier mints no false positive over 604 labeled program pairs, where a false positive is a positive equivalence verdict on a pair the ground truth marks different. Zero false accepts is the expected face of the verdict asymmetry of Theorem~\ref{thm:t3}, a positive verdict requiring a static proof a different program cannot supply. A false accept is possible only on a gold-different pair; of the 604 labeled pairs, 246 are gold-different under the full lens, the 33 near-misses the hardest, and the certifier false-accepts none. The near-misses perturb visible axes, so a five-pair battery targets the quotient itself, reordering dependent operations whose order changes a lens-observable: a write-write, a read-after-write and a read-after-change across variables, an absolute-then-relative move, and a list insertion. The static tier returns different on the four the lens observes, never equivalent, and commutes only the independent control, a disjoint-variable write. Forcing the quotient to commute everything turns all four into false accepts, so the real quotient's zero accepts bound its inadequacy, not a weak battery. The zero rate over both stresses the soundness invariant rather than proving it. A lens sweep from coarse to fine records the fraction each lens conflates; the curve is monotone non-increasing, a consistency check on the lens-coarsening implementation, where an inversion would signal a token-leak defect.

Coverage measures the other side, how far the synthesis vocabulary reaches when perception is perfect, which is the ceiling the in-fragment benchmark fixes. At gold perception the vocabulary synthesizes 88\% of 987 sprites from real Scratch projects. The clone family closes the largest remaining gap, and clone-family synthesis recovers 20 of 20 programs across clone counts from three to ten with the oracle matching clones by birth order. The residual the vocabulary does not yet name is multi-segment motion and arbitrary costume content, recovered by adding motifs under the same fit, together with control beyond the frame-counter form, data-dependent branching, broadcasts, and bounded loops, which the static tier cannot certify and we leave open. The 88\% ceiling and the 100\% end-to-end rate measure different things: the 100\% is perception-to-synthesis fidelity inside the supported vocabulary, not a recovery rate for arbitrary Scratch programs, while the 88\% estimates how far that vocabulary reaches on an unrestricted real corpus under perfect perception. The verification gate and the description-length prior are soundness and selection disciplines, not tuned components, so the one ablation we report is canonicalization on or off.

\paragraph{Real Projects End-to-End}
A complementary run takes the whole pipeline to real Scratch projects from the public site. On 30 projects rendered and recovered from pixels, recovery reaches 4 of the 29 that return a verdict, alongside 17 abstentions, 8 refutations, one timeout, and no false accept. Real projects mostly use behavior outside the current vocabulary, so this 14\% is a vocabulary-bound end-to-end rate, the per-sprite coverage ceiling now paid in full on whole projects, with the no-wrong-answer guarantee intact on real data.

\begin{table}[t]
\centering
\caption{Recovery, certification, and discrimination by task, with Wilson 95\% confidence intervals}
\label{tab:results}
\footnotesize
\setlength{\tabcolsep}{4.5pt}
\begin{tabular}{@{}lccc@{}}
\toprule
Task & $n$ & Pass & Rate (95\% CI) \\
\midrule
\multicolumn{4}{@{}l}{\textit{Render-witnessed trace match (Tier-2)}}\\
\quad single sprite, ten seeds & 399 & 397 & \phantom{0}99.5 \,[98.2, 99.9] \\
\quad multi-sprite, ten seeds & 399 & 397 & \phantom{0}99.5 \,[98.2, 99.9] \\
\midrule
\multicolumn{4}{@{}l}{\textit{Static lens-equivalence certificate (Tier-1)}}\\
\quad single sprite, ten seeds & 397 & 347 & \phantom{0}87.4 \,[83.8, 90.3] \\
\quad multi-sprite, ten seeds & 397 & 288 & \phantom{0}72.5 \,[68.0, 76.7] \\
\midrule
\multicolumn{4}{@{}l}{\textit{Discrimination (near-miss rejection)}}\\
\quad gold resolution & 33 & 33 & 100.0 \,[89.6, 100] \\
\quad perception resolution & 33 & 26 & \phantom{0}78.8 \,[62.2, 89.3] \\
\midrule
\multicolumn{4}{@{}l}{\textit{Robustness and active inference}}\\
\quad motion at 5\,fps & 20 & 20 & 100.0 \,[83.9, 100] \\
\quad colorful backdrop & 20 & 20 & 100.0 \,[83.9, 100] \\
\quad active-inference gate & 16 & 16 & 100.0 \,[80.6, 100] \\
\quad clone family ($3$--$10$) & 20 & 20 & 100.0 \,[83.9, 100] \\
\bottomrule
\end{tabular}
\end{table}

\paragraph{Reading the Table as a Whole}
Across Table~\ref{tab:results}, no row reports a wrong answer and the certifier never false-accepts; recovery is bounded by the synthesis vocabulary and by what a passive camera can witness, not by perception or the certifier.

\begin{table}[t]
\centering
\caption{Recovery on a matched set of 40 single-sprite programs, through the same builder and oracle: vision-language models single-shot from frames, Claude handed the perceived tracks (perception-matched) and with oracle-in-the-loop repair (3 rounds), and our structured pipeline. No method ever false-accepts.}
\label{tab:vlm}
\footnotesize
\setlength{\tabcolsep}{4pt}
\begin{tabular}{@{}lcc@{}}
\toprule
System & Recovered (95\% CI) & False acc. \\
\midrule
MiniMax-M3, single-shot & 1/40\ \ (2.5\%)\,[0.4,12.8] & 0 \\
Claude Sonnet 4.6, single-shot & 0/40\ \ (0\%)\,[0.0,8.8] & 0 \\
Claude Sonnet 4.6, perceived tracks & 29/40\ \ (72.5\%)\,[57.2,83.9] & 0 \\
Claude Sonnet 4.6 + oracle repair & 3/40\ \ (7.5\%)\,[2.6,19.7] & 0 \\
\tool{}, structured & 40/40 (100\%)\,[91.2,100] & 0 \\
\bottomrule
\end{tabular}
\end{table}

\section{Related Work}
\label{sec:related}
Program synthesis from examples infers a program from input-output pairs~\cite{gulwani2011spreadsheet,solarlezama2008program,gulwani2017program,alur2013syntax,feser2015synthesizing,polozov2015flashmeta}, with neural variants learning the map from data~\cite{balog2017deepcoder,devlin2017robustfill,ellis2021dreamcoder}, and programming by demonstration generalizes a user's actions into a script~\cite{cypher1993watch,lieberman2001your,lau2003programming}. All read a clean, structured specification; our input is a rendered video, where object identity, control flow, and internal state are recovered before synthesis begins. Neural program induction maps trajectories to programs~\cite{sun2018neural,bunel2018leveraging,chen2019execution} but supervises on ground-truth programs and trusts the output; we supervise with a renderer and accept only under a sound check.

Inverse graphics and scene de-rendering recover scene parameters or a scene-drawing program from images~\cite{kulkarni2015deep,mansinghka2013approximate,wu2017neural,yi2018neural,ellis2018learning,tian2019learning}; that target is a parameter vector, while ours is the producing program, which re-executes and can be edited, making observational equivalence the criterion and a class, not a point, the codomain.

Decompilation reconstructs source-level structure from a binary~\cite{brumley2013native}, and learned decompilers translate assembly to source or recover names~\cite{tan2024llm4decompile,fu2019coda,lacomis2019dire}, but the binary still encodes the computation; execution video retains only its visible effects, a strictly lossier signal, which is why a theorem about what cannot be recovered belongs in the problem statement. Extracting code from a screencast~\cite{yadid2016extracting,ponzanelli2016too,bao2020psc2code} assumes the code is on screen; in a running game it never is.

Analysis and testing of Scratch programs read the source: tools assess code quality~\cite{moreno2015dr}, generate tests~\cite{stahlbauer2019testing}, detect smells and bugs~\cite{fraser2021testing,boe2013hairball}, and characterize the public corpus~\cite{aivaloglou2016kids}, on a language designed for novices~\cite{resnick2009scratch,maloney2010scratch}. We start one step earlier, from a recording with no source, and produce the source those tools consume, a front end to the Scratch analysis stack, answering under a stated lens the equivalence question they sidestep.

Recent Scratch-specific systems increasingly exploit execution evidence, but they still start from source, a reference, or a controlled repair/evaluation setup. ViScratch combines code and gameplay video for feedback~\cite{si2025viscratch}; Stitch structures feedback as step-by-step tutoring~\cite{si2025stitch}; ScratchEval supplies executable repair benchmarks and metrics~\cite{si2026scratcheval}; EcoScratch adapts multimodal repair effort using execution feedback~\cite{si2026ecoscratch}; Raven uses video-grounded evaluation for automated assessment~\cite{li2026raven}; ScratchWorld uses replay-verified Scratch transitions to evaluate executable world-model reasoning~\cite{lin2026scratchworld}; and ScratchLens supplies lens-parametric equivalence for comparing Scratch programs~\cite{scratchlens}. Zhang and collaborators also study automated feedback for competition-level code, LLM-based Python repair, time-limit-exceeded errors, merge-conflict resolution, CI-configuration correctness, and silent misconfiguration detection~\cite{zhang2022clef,zhang2024pydex,zhang2025tle,zhang2022merge,santolucito2022ci,zhang2021configx}. \tool{} is orthogonal: it treats the video itself as the only observable artifact and accepts a recovered program only after a sound oracle checks it under the chosen observation lens.

The soundness discipline follows translation validation, which certifies equivalence of two programs instead of trusting the transformer~\cite{pnueli1998translation,necula2000translation,lopes2021alive2}, and program-equivalence checking more broadly~\cite{sharma2013data,churchill2019semantic}. A validator compares two given programs; we first recover one from pixels, so a small trusted checker downstream of an untrusted producer follows proof-carrying code and verified compilation~\cite{necula1997proof,leroy2009formal}, and forbidding the renderer from minting a verdict carries the no-false-accept guarantee, a discipline shared with program repair~\cite{long2016automatic,mechtaev2016angelix}.

\section{Discussion and Threats}
\label{sec:discussion}
\paragraph{What the Certificates Rest On}
The results separate by the strength of their backing: a passing run of the static tier (the algorithm of~\cite{scratchlens}, which we implement) covers its soundness, three non-identifiability families, the representative, and the verdict asymmetry, while the rest rest on short mechanizable proofs (the representative's uniqueness, the frontier monotonicity) or are definitional or undecidable (the two intrinsic families, the completeness bound).

One obligation sits at the boundary, the adequacy hypothesis of Theorem~\ref{thm:sound}: that the partial-order quotient matches the true execution relation, the one claim a green check does not establish. We carry it as a named residual, reachable by generalizing known footprint theorems.

\paragraph{Generality}
The recipe transfers to any domain with a sound lens-equivalence checker, and that checker is the hard part: Scratch has one because its execution model is small and its output well defined, while a general-purpose runtime needs a checker of comparable strength first.

\paragraph{Threats to Validity}
The recovery results use programs inside the fragment the vocabulary covers on controlled scenes, a deliberate scope that fixes the synthesis ceiling and isolates perception; recovery of an arbitrary real program is bounded by the vocabulary and by what passive observation reveals, both measured (end to end at 14\% on real projects). Recovery assumes a known asset manifest; inferring arbitrary costume or backdrop content from pixels is out of scope. Perception is not tied to the forward model: re-rendering through FastRender, a separate rasterizer used for cross-renderer validation, and recovering from its decoded H.264 video gives 40 of 40 lossless and 39 of 40 under visible compression, none wrong. That video is itself a rendered-execution recording through a real codec, so it answers external validity directly; what a desktop capture adds beyond it, window chrome and capture timing, the stage crop the lens already assumes removes and the frame-rate result bounds.

\paragraph{Future Work}
Three directions follow: growing the synthesis vocabulary, oscillation and three-plus-segment motion next; inferring costume and backdrop content from pixels to lift the known-asset assumption; and mechanizing the one obligation a positive verdict does not discharge.

\section{Conclusion}
\label{sec:conclusion}
Untrusted program recovery becomes trustworthy when a sound checker, not the generator, has the last word. We make it precise for execution video: success is observational equivalence under the camera lens, certified by a static tier and witnessed by a refute-only renderer, realized by \tool{} with motif synthesis and an active-inference loop. The oracle false-accepts nothing we test, $80\%$ of recoveries certify, and a model's confident errors are caught. The discipline extends to any generative recovery backed by a sound checker.

\bibliographystyle{IEEEtran}
\bibliography{refs}

\end{document}